# Considerations for Electromagnetic Simulations for a Quantitative Correlation of Optical Spectroscopy and Electron Tomography of Plasmonic Nanoparticles

Mees Dieperink[1], Alexander Skorikov[2], Nathalie Claes[3], Sara Bals[3], and Wiebke Albrecht[1]

[1]*Center for Nanophotonics, AMOLF, Science Park 104, 1098 XG Amsterdam, The Netherlands*
[2]*Computational Imaging Group, Centrum Wiskunde & Informatica, 1098 XG Amsterdam, The Netherlands*
[3]*EMAT and NANOlab Center of Excellence, University of Antwerp, Groenenborgerlaan 171, B-2020Antwerp, Belgium*
E-mail: w.albrecht@amolf.nl

## Abstract

The optical cross sections of plasmonic nanoparticles are intricately linked to the morphology of the particle. If this connection can be made accurately enough, it would become possible to determine a particle's shape solely from its measured optical cross sections. For that, electromagnetic simulations can be used to bridge the morphology and optical properties assuming that they can be performed in an accurate manner. In this paper, we study key factors that influence the accuracy of electromagnetic simulations. First, we compare several standard electromagnetic simulation methods and discuss in detail the effects of the meshing accuracy, choice of dielectric function and inclusion of a substrate for the boundary element method. To help the boundary element method's complex parametrization, we develop a workflow including reconstruction, meshing and mesh simplification steps to be able to use electron tomography data as input for these simulations. In particular, we analyze how the choice of reconstruction algorithm and the intricacies of image segmentation influence the simulated optical cross sections and correlate it to induced shape errors, which can be minimized in the data processing pipeline. In our case, optimal results could be obtained by using the Total Variation Minimization (TVM) reconstruction method in combination with Otsu thresholding and slight smoothing, which was important to create a reliable and watertight surface mesh using the marching cubes algorithm, especially for more complex shapes.

## Introduction

Metal nanoparticles (MNPs) can be exploited in a large range of optical applications [1], such as optical data storage [2], [3], sensing [4], [5], or photocatalysis [6], [7], [8], [9], [10], [11] owing to their highly tuneable localized surface plasmon resonances (LSPRs) ranging from the UV to the IR region. Next to the plasmonic and surrounding material, the morphology of the MNP is the key ingredient in defining the optical response [12]. For that reason, a lot of effort has gone into developing new protocols for the colloidal synthesis of MNPs with varying shapes and compositions [13], [14]. In particular, for gold NPs an amazing control over the morphology has been achieved and highly anisotropic NPs can now be routinely made [15], [16]. Examples include but are not limited to platelets [17], platonic solids [18], stars [19], [20], and even twisted [21] or wrinkled NPs [22], [23], [24]. The more complex the MNP shape, the less straightforward it becomes to correlate morphological and optical properties as several LSPR modes emerge. In addition, the polydispersity of the sample is often increasing for more complex shapes. An increased polydispersity leads to broader ensemble spectra, therefore possibly masking correlative features [25].

Consequently, analysis on a level of single particles has become increasingly important in understanding the structure-property correlation [9], [26], [27], [28], [29]. For plasmonic MNPs, the optical properties of single entities are mostly measured by scattering techniques such as dark-field scattering spectroscopy [30], [31], [32]. For a full picture, such optical data are then ideally correlated to morphological information on the same NP [33], [34], [35], [36], [37], [38], [39], [40]. Due to the small dimensions of typical plasmonic MNPs, electron microscopy cannot be avoided to obtain the necessary morphological information. Because of a simpler sample preparation and measurement workflow, scanning electron microscopy (SEM) is often preferred over transmission electron microscopy (TEM). In both cases, conventional SEM or TEM imaging provides 2D impressions of the 3D NPs. For the emerging increasingly more complex morphologies, however, 2D information is not sufficient and electron tomography (ET) has been established as a powerful technique to visualize but also quantify structural and morphological properties of MNPs [41], [42], [43], [44], [45], [46], [47], [48]. For crystalline materials, high-angle annular dark-field scanning TEM (HAADF–STEM) imaging is typically applied as the resulting signal satisfies the projection requirement for tomographic reconstruction [49]. For ET, the holder containing the TEM grid with the NPs is tilted over the maximal possible range, in practice often limited to around ±75° because of shadowing by the sample holder [50]. At every tilt angle a 2D projection image is acquired and finally all projection images are combined to retrieve the 3D morphology using a reconstruction algorithm.

Unfortunately, involving ET makes the correlative single particle workflow even more complex. For example, thin carbon-based TEM grids are well suited for ET but are ill-suited for optical scattering experiments. Using $SiO_2$ TEM grids with a few tens of nanometers in thickness can result in good optical data [40], [51], but these grids are fragile in handling, non-conductive and can lead to charging artefacts and shadowing at high tilt angles in the TEM thereby limiting the available angular range further. A new leverage for addressing this dilemma in correlating optical and structural properties of nanoparticles can be gained by employing electromagnetic simulations. On the one hand, they can provide the optical response based on the morphological input, e.g. obtained by ET. In this manner, next to the far-field response, the near-field can be determined as well, which is often the property of interest for plasmonic applications [4], [5], [6], [9], [52], [53]. On the other hand, if performed accurately enough, electromagnetic simulations can help us to do the inverse: getting information on the morphology from optical scattering data, e.g. via machine learning approaches [54]. A particular strength of such an optics-based approach is that it can provide morphological information about nanoparticles exposed to various conditions, such as high temperatures and liquid or gaseous environments typical in catalysis applications, for example. Such environments are difficult to introduce in electron microscopes [42], which limits our knowledge of realistic particle morphologies in *operando* conditions. Obtaining such insight from optical data would therefore be invaluable in nanoplasmonics.

However, performing accurate electromagnetic simulations based on ET input is surprisingly non-straightforward as several factors need to be taken into [55]. First, several different electromagnetic simulation methods exist, each with their own advantages and disadvantages [35], [56], [57], [58]. Second, for each method different parameters influence the convergence results. In addition, for plasmonic simulations, the dielectric function of the metal and simulation of the accurate surrounding are critical [59], [60], [61], [62], [63]. Third, ET reconstructions need to be segmented and possibly surface-meshed to be useable as input for such simulations. To do so, a variety of different reconstruction [64], segmentation [65] and meshing algorithms exist [66]. So far, no comprehensive study exists that compares all these factors and the resulting parameter space quantitatively.

In this study, we tackle such a quantitative comparison. After weighing several standard electromagnetic simulation methods against each other, we discuss the effects of the main parameters for the boundary element method (BEM). Finally, we carefully design a workflow to be able to use electron tomography data as input for our simulations. We focus the main discussion on Au nanorods but compare the workflow results on different shapes as well. We first discuss the effects of the meshing accuracy, choice of dielectric function, and the inclusion of a substrate. Then, we go into detail for our case study of a Au nanorod, in which we highlight the effect of the choice of reconstruction algorithm, and the intricacies of image segmentation. Ultimately, we compare how morphological changes, induced by different processing pipelines, influence the results of the BEM simulations.

# Results and Discussion

<u>Considerations for electromagnetic simulations</u>

We started by choosing a suitable electromagnetic simulation method. For efficient characterization of nanoparticles with complex geometries, we need fast and accurate electromagnetic simulations. To discover what method is best for our purpose, we performed careful convergence testing for the Discrete Dipole Approximation (DDA), the Finite-Difference Time-Domain (FDTD) method, the Discontinuous Galerkin Time-Domain (DGTD) method, and BEM. A description of the methods and computational details can be found in section S2.1 of the Supporting Information (SI). These classes of methods were chosen because they are the most widely used methods for simulating the optical properties of metal nanoparticles, with each of them representing a different way of morphology discretization [35], [56], [57], [58]. It should be noted that DGTD represents a class of finite element method solvers, with COMSOL and CST being alternative commercial implementations. We applied the four different methods for a spherical Au nanoparticle, because the simulated cross sections can be quantitatively compared to the accurate analytical Mie solution [67]. However, for a fair comparison it is important to make sure that the different simulations are performed with the optimal parameters ensuring a good convergence for each method[68]. Convergence is reached when changing a parameter $d$, which is relevant for the accuracy of the simulation method, e.g. the discrete element size, does not significantly change the result $\sigma$ of the simulation, e.g. the scattering cross section spectrum. The magnitude of change $\Delta\sigma$ can be measured by various metrics, for example normalized root-mean-square deviation and reflects the error due to discretization Equation 1.

$$\Delta\sigma(d_i) = \sqrt{\frac{\int \left(\sigma(d_{i,\lambda}) - \sigma(d_{i-1,\lambda})\right)^2 d\lambda}{\int \sigma^2(d_{i,\lambda}) d\lambda}}$$

(1)

where $d_i = d_1, \ldots, d_n$ is a monotonic sequence of simulation parameter values, $\sigma(d_{i,\lambda})$ is the wavelength-dependent simulation result, and $d\lambda$ is the wavelength step. Typically, the user optimizes the simulation by varying the value of $d_i$ until an acceptably small value of $\Delta\sigma$ (discretization error) is reached. The same equation can be used for calculating the error of the scattering spectrum compared to a known reference spectrum:

$$\Delta\sigma_{ref}(d_i) = \sqrt{\frac{\int \left(\sigma(d_{i,\lambda}) - \sigma_{ref}(\lambda)\right)^2 d\lambda}{\int \sigma_{ref}^2(\lambda) d\lambda}}$$

(2)

in which the exact error of the result $\sigma(d_{i,\lambda})$ can be calculated compared to a reference result $\sigma_{ref}(\lambda)$, e.g. the analytical Mie solution.

The convergence of a simulation method can be affected by multiple interdependent parameters. For instance, the simulation time and auto-shutoff parameter in FDTD strongly depend on each other. In case the simulation time is chosen too short, energy is still present in the simulation region when the simulation finishes, leading to a non-converged result. Controlling the end of the simulation with the auto-shutoff parameter results in converged simulations, but an inconsistent simulation time. Therefore, each relevant simulation parameter was optimized successively until the desired convergence threshold was reached for each of them. Details of the different parameters that were optimized for each method can be found in section S2.1 of the SI.

Figure 1A compares the resulting scattering spectra of a Au sphere with a diameter of 50 nm immersed in oil (n=1.51) for the four electromagnetic simulation methods to the analytical Mie solution. We limited ourselves to scattering since most experimental single particle setups measure scattering and not absorption. In order to quantitatively compare the different simulation methods, we used results obtained with a same discretization error of 2%, which was chosen as a trade-off between the accuracy and the simulation time. The zoomed inset shows that the cross section calculated by DGTD resembled Mie theory most. However, from the legend it becomes clear that it was not the fastest method, which instead was BEM. The reported times correspond to the simulation time it took to reach a discretization error of <2%, which resulted in a spectrum error of <5% compared to Mie theory for all methods (big circles in the green inset in Figure 1B).

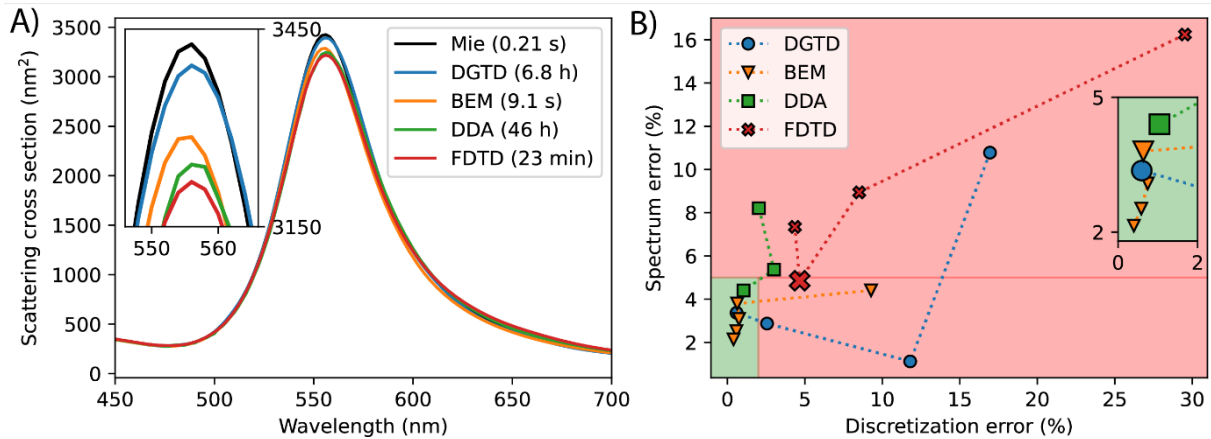

**Figure 1.** Comparison to Mie theory of simulated scattering spectra of a Au sphere with a diameter of 50 nm immersed in oil (n=1.51) using the Discontinuous Galerkin Time-Domain (DGTD) method, the Boundary Element Method (BEM), the Discrete Dipole Approximation (DDA), and the Finite-Difference Time-Domain (FDTD) method in a qualitative (A) and quantitative (B) manner. The times in the legend of (A) correspond to the times it took to produce the results in (B) that were the first in the optimization sweep to reach a discretization error of <2% and a spectrum error of <5% (within the green rectangle), except for FDTD which did not reach this error of discretization for the swept parameters. These displayed spectra in (A) are indicated with the enlarged markers in (B).

Figure 1B displays the results of the convergence tests which were performed by sweeping the following discretization parameter for the different methods: the maximum edge length of the tetrahedral elements for DGTD, the number of triangles for BEM, the number of dipoles for DDA, and the edge length of the cubical elements for FDTD. These parameters were defined as the most critical for the respective simulation methods (see S2.1 in SI). The discretization error was calculated by comparing each refinement step with the previous step using Equation 1. The errors of the obtained cross sections were calculated using Equation 2 with Mie theory as a reference. For instance, the rightmost point for DGTD was obtained by calculating the cross sections with edge length values of 15 and 10 nm. To calculate the discretization error, the result from 10 nm was taken as next step (i) in Equation 1, while the result from 15 nm was taken as initial step (i-1). To calculate the spectrum error, the result from 10 nm was taken as the result in Equation 2 and Mie theory was taken as the reference. This resulted in a discretization error of 17.0% and a spectrum error of 10.8%.

As expected, the general trend is that refinement of the mesh resulted in a lower discretization error and lower spectrum error compared to Mie theory. However, only BEM truly followed this trend in a straightforward manner. DDA showed an irregular trend in discretization error, while both FDTD and DGTD showed an irregular trend in spectrum error. For example, the resulting cross section from DGTD seemed to move away from Mie theory for a finer meshing parameter, while the convergence still went down. This makes these methods less predictable, as it is unsure if a lower discretization error (and hence a better convergence) also results in a lower spectrum error.

Compared to the other simulation methods, BEM was also orders of magnitudes faster. The enhanced simulation speed stems from the fundamental difference of BEM: Maxwell's equations only need to be solved at the surface of the nanoparticle and not for the whole volume as is the case for the other methods[58]. For FDTD, the fields additionally need to be propagated in a large region outside the particle. The spectrum error for BEM could be further reduced to 2.1% at the cost of a higher computation time (Figure 1B). BEM also converged fastest with respect to changing the meshing parameter. Although for DGTD the lowest spectrum error in Figure 1A was slightly smaller (1.1%), although at a higher discretization error (around 12% as shown in Figure 1B), the 2700 times longer simulation time and unpredictable convergence behaviour made us favour BEM over DGTD. It should be noted that Trügler et al. also reported the faster computation speed of BEM compared to other methods but did not report quantitative differences in scattering cross sections when comparing different simulation methods [56]. Moreover, in that study the normalized scattering spectra were compared, and no convergence testing was mentioned in the discussion on computation time. Since we are interested in simulating absolute scattering cross-section spectra, these small differences between methods become important. For our purposes, BEM delivered the best combination in terms of speed and accuracy. Therefore, we use BEM throughout the rest of the paper.

Now that the need for convergence testing is clear, it is key to look at the individual simulation parameters of BEM in more detail. In this paper we discuss the three parameters that influence the resulting cross sections most:

- Meshing of structure: Depending on the shape of the nanoparticle, the surface plasmon is localized around regions of high curvature or small gaps, e.g. at tips in nanorods. It is important to finely mesh these parts to get accurate results.
- Dielectric function of plasmonic material: Different experimentally determined dielectric functions yield significantly different results. We would like to advocate for better awareness in its choice.
- Substrate: The substrate is often excluded in electromagnetic simulations but cannot be neglected for accurate comparisons between simulations and experiments as in most single particle experiments the optical properties of the nanoparticles are measured on a substrate.

In the remaining part of the section we focus the discussion on the nanorod as the particle morphology because it is the most widely used plasmonic anisotropic nanoparticle shape [69]. For a nanorod, the electric field enhancement is highest at the tips, which needs to be taken into account when meshing its surface. In the MNPBEM toolbox that we use for BEM simulations, a nanorod is defined with the three parameters listed below (Figure 2A). We determined the corresponding conversion into a physical size by looking at the resulting sizes of the surface triangles for a given parameter [70]:

- The discretized polar component of a rod is denoted as $n_\varphi$ and the conversion of the corresponding discretized size into nm is given by: $dis_\varphi \approx \frac{\pi(d+1)}{n_\varphi}$ where $d$ is the diameter of the rod.
- The discretized azimuthal component of the hemispherical caps is denoted as $n_\theta$ and the conversion of the corresponding discretized size into nm is given by: $dis_\theta \approx \frac{d+1}{n_\theta}$
- The discretized meshing along the cylinder length is denoted as $n_z$ and the conversion of the corresponding size size into nm is given by: $dis_z \approx \frac{l-d+1}{n_z}$ where $l$ is the length of the rod.

We hope that these estimated physical sizes of the meshing elements can be useful when comparing different simulation methods. The provided analysis illustrates that it is important to optimize the number and distribution of triangles in the mesh to obtain accurate electromagnetic simulations in a realistically attainable time. For different shapes, the optimal values are expected to be different from this nanorod example, and we advise to perform the optimization described above to obtain accurate results.

By changing the number of each component, convergence tests were performed. Figure 2B shows relative differences in discretization error (Equation 1) between simulations that were performed with different combinations of the three discretization parameters $dis_\varphi$, $dis_\theta$, and $dis_z$ for a Au nanorod with a diameter of 30.0 nm and a length of 96.5 nm (AR=3.2). The sweep direction in this plot goes from high $dis_z$ to low $dis_z$ values, and by changing $dis_\varphi$ and $dis_\theta$ in parallel, the effect of all parameters is displayed at once. Here, a lower value and brighter colour indicates that the simulation has converged more, which is what we aim for. As was discussed for Figure 1, the convergence in BEM can be directly translated into the simulation accuracy and is therefore a good metric for the parameter sweep.

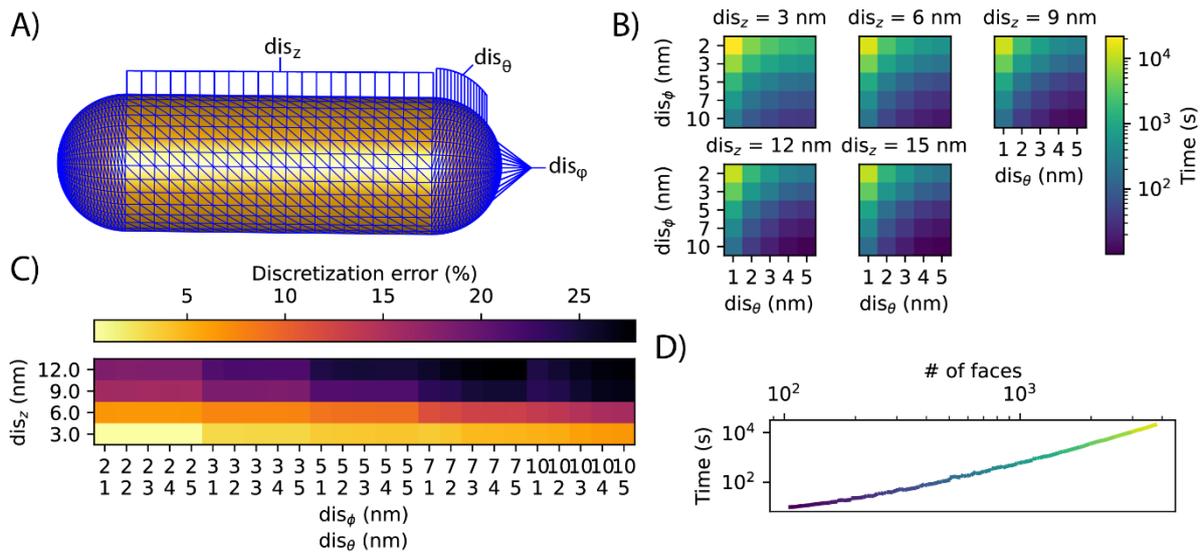

**Figure 2.** The effect of discretization for a Au nanorod with a diameter of 30.0 nm and a length of 96.5 nm (AR=3.2). (A) Visualization of the surface mesh where the discretization values are $dis_z \approx$ 3 nm, $dis_\theta \approx 1$ nm, and $dis_\varphi \approx 3$ nm. (B) Relative differences in discretization error and corresponding simulation times as a function of (C) different discretization parameters and (D) the total number of faces.

The general message from Figure 2B is that changing $dis_z$ influenced the discretization error most, but without small values for $dis_\varphi$ and $dis_\theta$ the discretization error did not reach an acceptable level. For instance, when we decreased $dis_z$ from 12 to 3 nm, while keeping $dis_\varphi \approx 10$ nm and $dis_\theta \approx 5$ nm, the discretization error dropped from 27.6% to 6.6%, which is still above our above defined threshold of 2%. Then, when we changed $dis_\varphi$ from 10 to 2 nm, we obtained a discretization error of 0.3%, which is well below our above defined threshold of 2%. It might be surprising that $dis_z$ influenced the discretization error most although it is connected to the least curved part of the particle. We believe that one explanation might be that a large difference between $dis_z$ and $dis_\varphi$ leads to highly non-equilateral meshing triangles, which are known to be detrimental for finite element simulations [71], [72].

The simulation time for the meshing with the smallest $dis_z$ was already 25 minutes (Figure 2C), which was significantly longer than for the more simple spherical geometry discussed in Figure 1. Figure 2C demonstrates that the variation of meshing parameters had a strongly non-linear effect on the simulation time. Therefore, for rods and other anisotropic shapes, the balance between accuracy and speed needs to be adjusted. By allowing a discretization error of 3.9%, for example, the simulation time could be decreased to 2.6 min for $dis_z \approx 3$ nm, $dis_\varphi \approx 5$ nm, and $dis_\theta \approx 3$ nm. The simulation time is directly linked to the total number of faces (Figure 2D), which can be estimated according to Equation 3. It should be noted, that this an empirical estimate and not derived from mathematical arguments. However, it is a helpful estimate when deciding on the meshing accuracy at least in the case of nanorods. From Equation 3 it can be seen, that $dis_z$ should influence the simulation time least and this is indeed observed in Figure 2C.

$$n_{faces} = n_\varphi(n_z + 2n_\theta)$$

(3)

From Figure 2 we can conclude that the following approach should be followed for a nanorod. A low value for $dis_z$ is ideally chosen to reach a low base discretization error, like 3 nm for this nanorod. Luckily, this can be achieved without paying a high penalty in simulation time. It should be kept in mind to avoid large distortions of the triangles by choosing similar values for $dis_z$ compared to $dis_\varphi$. To obtain an even lower discretization error, fine meshing with $dis_\varphi$ and $dis_\theta$ is required, on the order of 1 nm for both for our specific nanorod. However, these parameters heavily affect the simulation time as a low $dis_n$ leads to higher number of faces (Figure 2C and Equation 3) and a compromise in terms of size should be made for one of them. For high aspect ratio rods, the effect of $dis_\varphi$ on the simulation time is expected to increase even more since the parameter affects the meshing of the whole rod because it gets multiplied with $dis_z$, which in turn needs to be more finely meshed to not distort the triangles. Equation 3 and Figure 2D can help to estimate what the expected simulation time is for a specific combination of parameters. To keep the comparison as general as possible between different rods the following parameters were used throughout the remainder of this section:

- $dis_\varphi \approx 3$ nm
- $dis_\theta \approx 1$ nm
- $dis_z \approx 3$ nm

Next to the meshing parameters, the choice of dielectric function of the plasmonic material influences the simulated scattering cross sections. In Figure 3A we compare BEM simulations with common choices of Au dielectric functions for a Au nanorod of the same size as used for Figure 2 and for the above determined mesh size parameters [73], [74], [75], [76]. The peak position and height of the LSPR obviously strongly depended on the chosen dielectric function. For example, for the simulations using the dielectric function from the single crystalline, but rough sample from Olmon et al. [73] and the polycrystalline, but smooth sample from McPeak et al. [76] the peak positions were 1.45 eV (855.1 nm) and 1.52 eV (816.7 nm), respectively, which is a relative difference of 4.5% (4.69%). The peak heights were 45946 nm$^2$ and 55940 nm$^2$, respectively, which is a relative difference of 17.9%. It is remarkable that these dramatic peak changes arose from relatively small differences in the $n$ and $k$ values (Figure 3B), which stem from the difference in the preparation protocol of the metal films. Two main parameters play an important role here: the crystal grain size and the surface roughness [76], [77]. The larger the grain size and the smaller the surface roughness, the better is the plasmonic performance. For the longitudinal plasmon of the studied Au NR a better film should therefore result in a higher energy resonance and larger cross section. When using the dielectric function for Au of McPeak et al., we obtained the highest cross section and highest energy LSPR. This is expected as the authors put in a lot of effort to optimize the optical performance of their Au films with large grain sizes and low surface roughness. For us it was surprising to see, however, that the single-crystalline film of Olmon et al. displayed a significantly red-shifted and lower scattering cross section. Although the grain sizes might be expected to be bigger, the surface roughness of their prepared films must have been larger than for McPeak et al. as also evidenced from their AFM data. In the end, the choice of dielectric function needs to be made on a case-by-case basis depending on the nanostructure preparation. Ideally, for our system we should use a dielectric function that is measured on a single particle, but, to our knowledge, this has only been done for a small wavelength range [78]. Therefore, we settle for the dielectric function measured by McPeak et al. It should be noted that other effects, such as surface damping need to be added to the dielectric function for small nanoparticles [39], [40], [79], [80]. However, for our 30 nm diameter nanorod surface damping is negligible. In addition, for nanoparticle sizes below 5 nm, quantum size effects need to be considered as well, which can also be incorporated into the dielectric function [81].

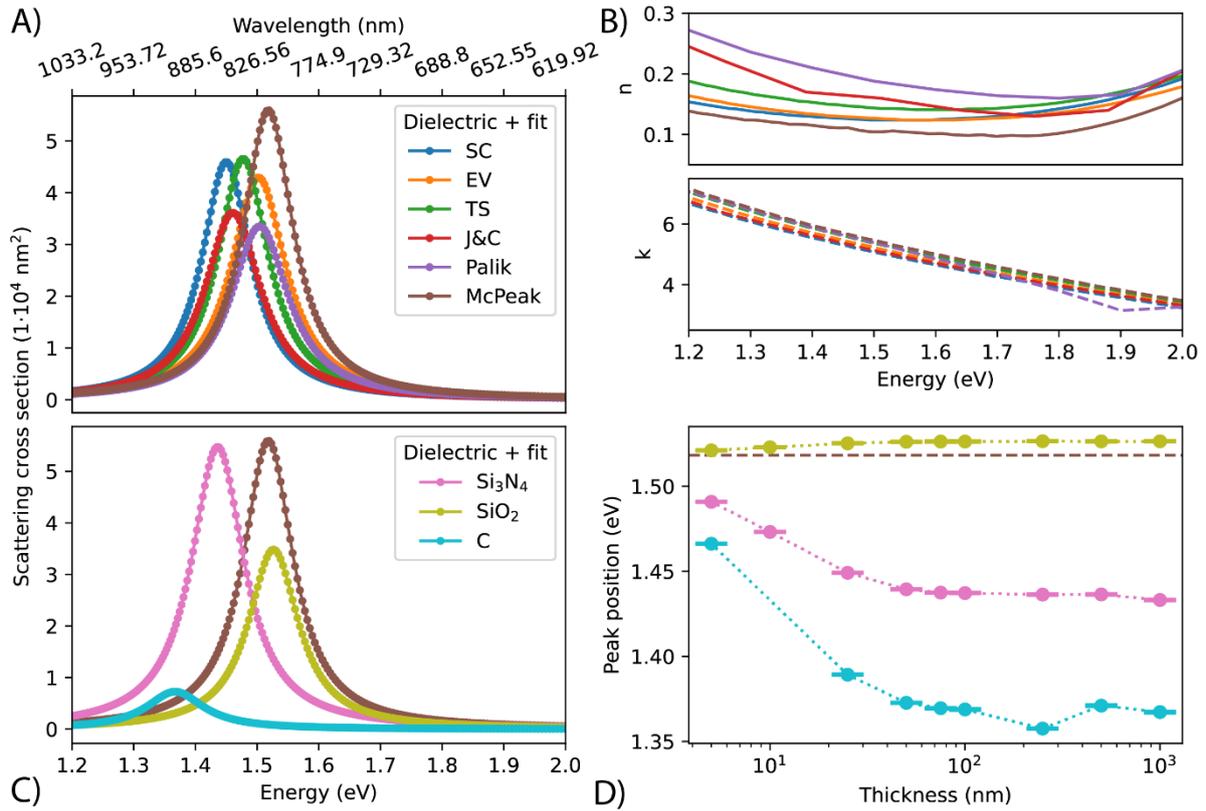

**Figure 3.** The effect of the dielectric function for a Au nanorod with a diameter of 30.0 nm and a length of 96.5 nm (AR=3.2), discretized with $dis_z \approx 3$ nm, $dis_1 \approx 5$ nm, and $dis_\varphi \approx 3$ nm and immersed in oil (n=1.51) (A and B), or on top of a substrate (C and D). (A) BEM scattering spectra simulations using the single-crystalline (SC), evaporated (EV), and template-stripped (TS) data from Olmon et al. [73], the data from Johnson and Christy [74], the data from Palik [75], and the data from McPeak et al. [76] with their corresponding $n$ and $k$ values shown in (B). (C) BEM scattering spectra simulations using the data from McPeak et al. for the nanorod on different substrates, surrounded by immersion oil (n=1.51), using the substrate dielectric constants from references [82], [83], [84] for an infinitely thick substrate. (D) The peak position of the LSPR for different substrate thicknesses using the system in (C) where the brown dashed line is the peak position of the simulation without a substrate.

For almost all single-particle optical experiments, such as in the commonly used dark-field scattering spectroscopy, a substrate is used on which the sample is deposited, often standard glass slides. However, for correlative studies on MNPs a substrate is required that can be used both for optical and electron tomography measurements. It therefore needs to be electron transparent and typical TEM substrate thicknesses are below 50 nm. In Figure 3C BEM simulations of the same Au nanorod including a substrate surrounded by immersion oil are shown for three common materials for TEM grids: $SiO_2$, $Si_3N_4$ and C [82], [83], [84]. The significant difference in material clearly affected the LSPR of the nanorod. For instance, the lossy nature and high refractive index of C (Figure S1) damped and red shifted the LSPR significantly. Finally, as expected [61], the substrate thickness mattered as well (Figure 3D). The LSPR shift with increasing substrate thickness was largest for C due to the largest dielectric constant discrepancy with respect to the surrounding oil. Since $SiO_2$ is much better index matched to the surrounding oil, the shift was marginal. To exclude thickness effects, due to e.g. locally varying thicknesses, $SiO_2$ TEM grids immersed in oil during the optical measurements are therefore ideal. When the optical measurements are done in non-index matched environments, the thickness of the underlying substrate needs to be clearly considered when performing quantitative electromagnetic simulations.

It should be noted that the choice of substrate, dielectric function and meshing accuracy does not only influence the far-field properties as highlighted here, but also need to be considered for near-field simulations. It should also be noted that our simulated gold nanorod is a spherocylinder and that synthesized crystalline nanorods contain crystal facets. Depending on the contact area on the substrate, this influences the optical cross sections as well [85], [86]. When comparing the simulated spectra to experimentally measured ones, this effect is automatically included with our workflow as the input shape is based on the experimentally measured morphology from tomography as detailed in the next section.

From electron tomography to mesh

When correlating optical and structural properties of single plasmonic nanoparticles, the most straightforward approach is to use 2D SEM or TEM images to retrieve the structure of the nanoparticle. As discussed in the introduction, this approach is not applicable to complex shapes, and one needs to resort to electron tomography. But even for seemingly symmetric particles extracting a 3D structure from 2D images might lead to incorrect estimation of morphological parameters, complicating the structure-property interpretation. An example is shown in Figure 4 for a Au nanorod imaged by HAADF-STEM (Figure 4A). In order to extract the length and width of the Au nanorod, the image needs to be segmented to differentiate the particle from the background signal. The corresponding pixel intensity histogram of the image in Figure 4A is displayed in green in Figure 4C, revealing two peaks in the intensity distribution. The peak at lower intensities corresponds to the background, while the peak at higher intensities corresponds to the foreground.

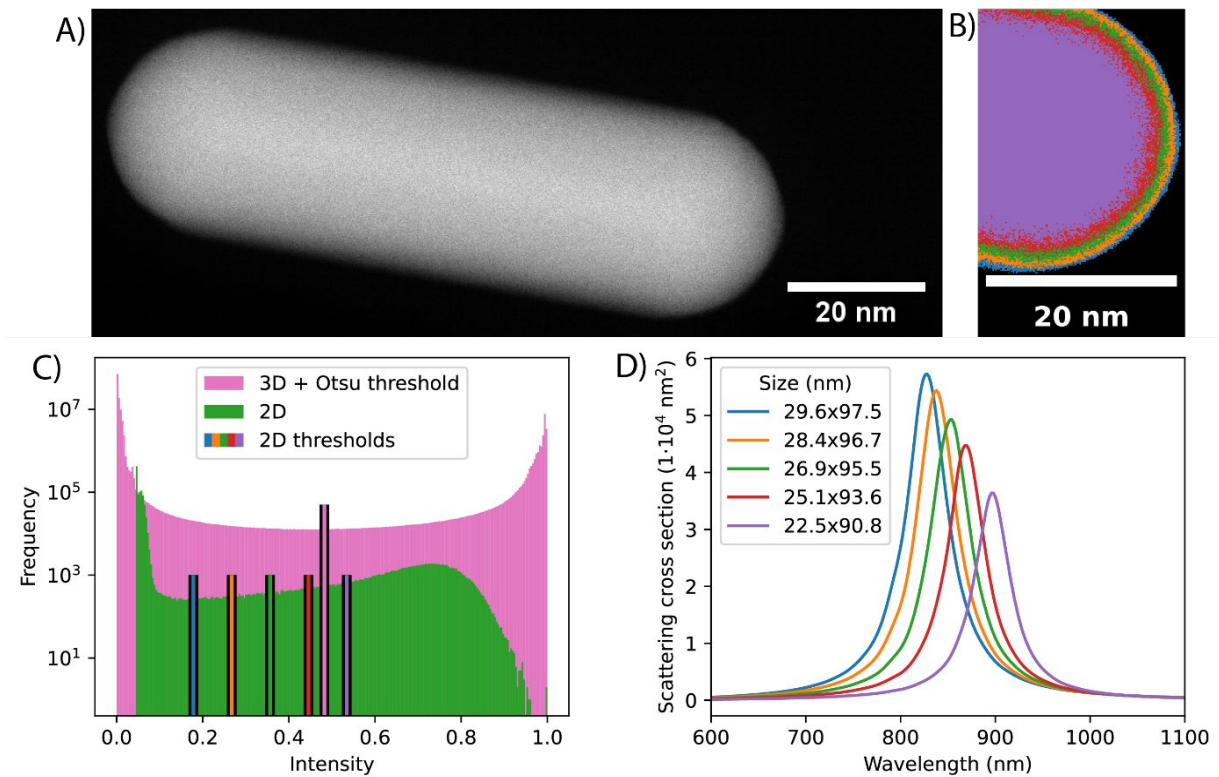

**Figure 4.** The effect of segmentation of high-angle annular dark-field scanning TEM (HAADF–STEM) projections and a reconstruction thereof. (A) HAADF-STEM image of a Au nanorod on a homemade holey-C Cu TEM grid at 0° tilt angle. (B) Different segmentations of the tip for which the colors correspond to (C) and (D). (C) Pixel intensity distributions with corresponding Otsu thresholds or fractions thereof (0.5: blue, 0.75: orange, 1: green, 1.25: red, 1.5: purple) of reconstructed data from the complete 3D data set (pink) and the 2D data shown in (A) (green). (D) Simulated BEM scattering spectra of Au nanorods with sizes corresponding to the legend, resulting from the different segmentation thresholds in (B) and our 2D fitting algorithm (S2.2.5), discretized with $dis_z \approx 3$ nm, $dis_1 \approx 5$ nm, and $dis_\varphi \approx 3$ nm and immersed in oil (n=1.51) using the gold dielectric function measured by McPeak et al. [76].

One problem for 2D data like the one in Figure 4A is the rather smooth transition between the two distributions due to the thickness-dependent HAADF–STEM intensity and the strong influence of Poisson noise, which makes it difficult to segment the particle [87]. Several algorithms exist for segmentation, with the most common one being Otsu's method [88], [89], but none of them are designed for the case of a smooth transition between two intensity distributions. Therefore, choosing a threshold value in this is not straightforward as demonstrated in Figure 4B,C. The vertical lines in Figure 4C correspond to threshold values using Otsu's method (green line) or fractions thereof (0.5: blue, 0.75: orange, 1.25: red, 1.5: purple). All these values could seem like a reasonable choice for separating the two distributions. However, the effect of this choice can be seen directly in Figure 4B, where the colors correspond to the threshold choice. In combination with the finite pixel size, the choice of threshold for the 2D HAADF–STEM image resulted in highly changing extracted dimensions of the nanorod. The changing nanorod sizes directly influenced the simulated LSPR of the MNP tremendously, which is shown in Figure 4D by performing BEM simulations with models of fitted sizes using the 2D fitting algorithm described in section S2.2.5 of the SI. This illustrates the uncertainty of relying on 2D data when aiming for accurate simulations of electromagnetic properties of plasmonic nanoparticles.

We proceeded by comparing the 2D pixel intensity histograms to the voxel intensity histogram of the 3D data set. The latter was obtained by acquiring a set of 2D projection images in the tilt range of -77° to +72° (details in section S1.2 of the SI), which were subsequently reconstructed using the total variation minimization (TVM) algorithm. The influence of the choice of reconstruction algorithm and segmentation method will be detailed later. For now, the pink histogram in Figure 4D demonstrates another advantage of using ET in addition to providing the realistic 3D morphology: The separation of the background and foreground became significantly clearer after reconstructing an experimental tilt series of the Au nanorod, reducing the uncertainty in the segmentation process. Segmenting and fitting the 3D reconstruction of our experimental example resulted in a diameter of 30.0 nm and a length of 96.5 nm (AR=3.2). It should be noted that fitting the nanorod shape to the 3D data was done for the sake of comparing the sizes to the 2D results. However, in the following we use the 3D output of the tomographic reconstruction directly. This approach becomes particularly important for simulations of MNPs with complex geometries, where it is not possible to guess a 3D shape from 2D images. To make the output of the tomographic reconstruction suitable for BEM simulations, the voxelized reconstruction needs to be transformed into a triangular surface mesh requiring segmentation of the particle as an intermediate step. In the following, we discuss considerations for an optimized workflow to achieve this.

To evaluate the importance of the possible errors that are introduced during the different steps along the way, we used a well-defined ground truth. For that, we simulated electron tomography data for a nanorod using the sizes from the fit to the experimental 3D data from Figure 4. Since electron microscopy data contains noise and image artefacts, we need to account for this when simulating the 2D projection images. The most prominent contributions to this are Gaussian blurring caused by defocus and astigmatism, and Poisson noise arising from the discrete nature of the recorded signal [87]. Additionally, the background signal from the sample support needs to be taken into account for a realistic representation of STEM images. The STEM images were simulated by forward projecting a voxelized model of a nanorod with the fitted sizes using the ASTRA toolbox 2.1.0 for the experimental tilt angles (details in S1.2) [90]. Then, for each 2D projection a Gaussian filter was applied to model blurring, followed by simulating the background signal from the sample support. To stay as close to experimental parameters as possible, we modeled the relative background signal level by calculating the mean of the background values for every experimental projection image from the Au nanorod from Figure 4. We did that by first removing the particle from the 2D image through segmentation using a threshold that made sure that the whole particle was removed and calculating the mean of the remaining image. Figure 5A shows that the background level increased with increasing tilt angle because of carbon contamination deposition throughout the experiment, which can be clearly observed when comparing a 2D HAADF–STEM image taken before and after the tilt series (see insets). The sharp increase at the first and last tilt angle can be attributed to detector shadowing. The estimated background level was added to each simulated projection image independently, and Poisson noise was applied on a pixel-by-pixel basis after manually tuning the scaling of the signal for the particle and the background to match the experimental noise levels. To assess the result of noise addition to the simulated data, line profiles were compared between simulated and experimental data. The insets in Figure 5B show representative projections from both experimental and simulated electron tomography data and the extracted line profiles show a good qualitative match.

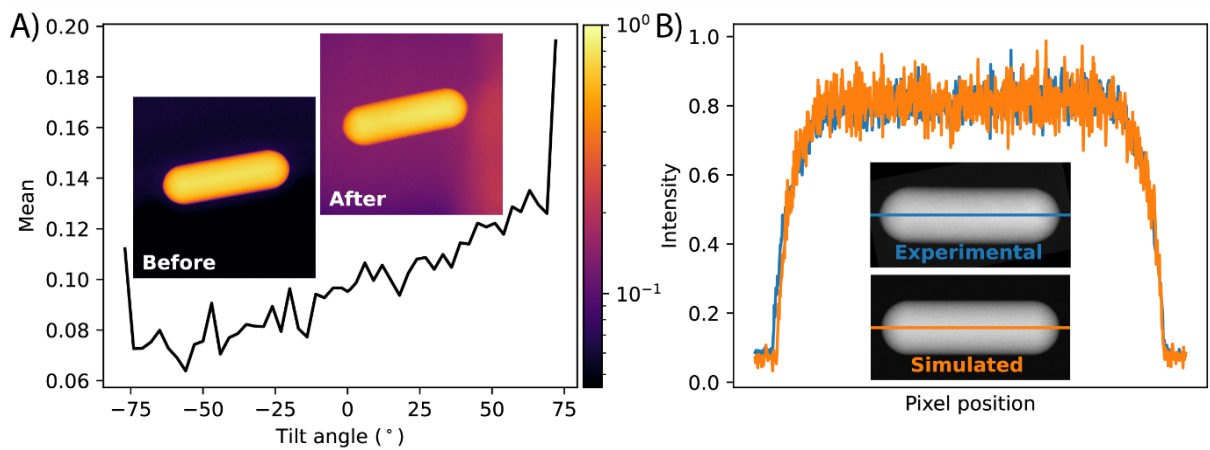

**Figure 5.** Simulating electron tomography data with realistic experimental input. (A) shows the mean values of the background of the data from Figure 4 at different tilt angles. The inset shows a projection image at 0° before and after tilting with a logarithmic intensity scale to show the background more clearly. (B) compares representative line profiles of experimental (blue) and simulated (orange) 2D projection data. The inset shows the corresponding projection images and lines along which the profiles were extracted.

Using the tomography data simulated for the ground truth shape, it becomes possible to compare different methods for the different steps in our processing pipeline. For a quantitative comparison we used the ground truth shape on a voxel grid with the same voxel size as used in the experiments as a reference for calculating the shape error $E_S$ induced by the different choices in the data processing steps:

$$E_S = \frac{\sum |Vox_{sim} - Vox_{ref}|}{\sum |Vox_{ref}|} \cdot 100\%$$

(4)

The shape error takes into account misclassified voxels and therefore reveals shape deviations and possible misalignments of two structures even if the total volume is the same. In the case of our simulated data, there is no effect of misalignment but it needs to be kept in mind when using experimental data, for which careful object registration needs to be performed first for an accurate shape error calculation [45]. To calculate the shape errors, the reconstructions needed to be segmented. For the comparison here, we used the Minimum method [65], which outperformed other methods for the nanorod shape as detailed below. A more in-depth discussion of the segmentation method is found provided in section S2.2.3 of the SI with Figures S2 and S3.

As a first step, we compared different pre-processing methods applied to the input projection images before performing tomographic reconstruction. It was observed that pre-processing had negligible influence on the final result due to the high signal-to-noise ratios for our data (details in section S2.2.4 of the SI). Interestingly, smoothing the input projection images resulted in the same marginally improved shape error as smoothing the 3D reconstruction as a whole in the case of a low noise reconstruction (see Figure S5 in section S2.2.4 of the SI). Therefore, in the following discussion we used the unprocessed projection data. Next, three common iterative tomographic reconstruction algorithms were compared: expectation-maximization (EM) [91], simultaneous iterative reconstruction technique (SIRT) and total variation minimization (TVM) [64]. These algorithms utilize different assumptions about the reconstructed object: EM and SIRT algorithms produce maximum likelihood reconstructions in case the input data are coming from Poisson or normal distributions, respectively. Both of these algorithms minimize the discrepancy between the input data and the projection of the reconstructed object, and TVM incorporates an additional objective of minimizing intensity variations in the solution, thereby promoting smooth, piecewise-constant reconstructions.

Figure 6 shows the obtained shape errors with reference to the voxelized ground truth for a variety of different reconstructions for which the reconstruction method, the number of algorithm iterations, the object shape, and the angular sampling range were varied. Figure 6A displays the effect of the number of iterations, illustrated using the EM algorithm. Increasing the number of iterations from 15 to 25 decreased the shape error but more iterations led to its increase. This effect is common to the iterative algorithms that minimize the discrepancy between the reconstruction and the noisy input data. At lower iterations the algorithm converges to the true solution, but eventually the reconstruction is overfitted to experimental noise, and the error compared to the ground truth increases. Thereby, there is an optimal number of iterations depending on the noise level in the input data, which in our case was around 25 iterations [91].

The effect of the limited angular sampling range was evaluated by comparing the shape errors for the reconstructions obtained for different tilt ranges with 25 iterations of EM (Figure 6B). Non-surprisingly, the shape error increased when the number of tilt angles decreased. The 'residual' error of 1.18% for a full tilt range of ±90° is a combination of the discrete tilt step and segmentation process [50]. For the experimental tilt range of -77° to +72° the shape error increased marginally to 1.63%. However, when the tilt range was significantly decreased to ±45° the shape error severely increased to 7.26%.

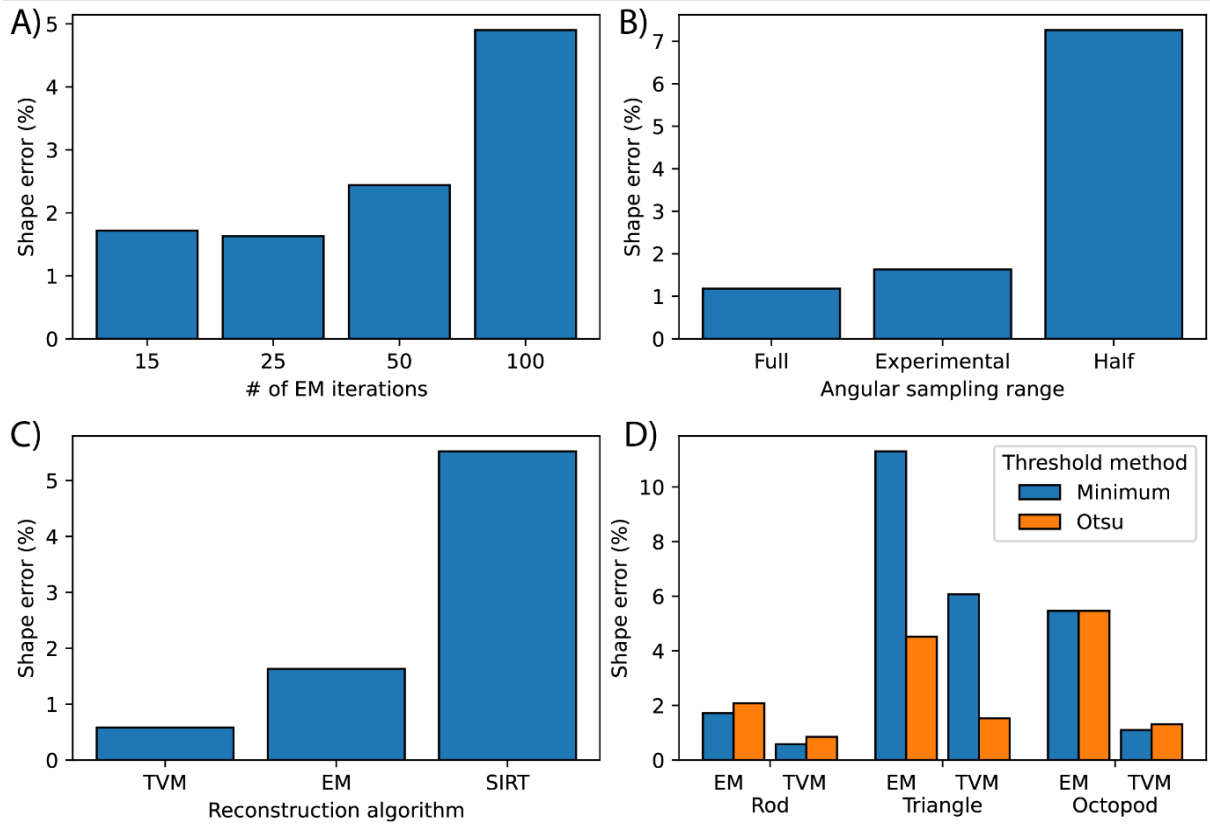

**Figure 6.** The effect of the number of iterations (A), the angular sampling range (B), and the reconstruction algorithm on the shape error (C) when reconstructing our simulated data from Figure 5B. (D) Comparison of EM and TVM for three different shapes using the Minimum and Otsu segmentation methods.

When comparing different reconstruction algorithms, EM significantly outperformed the more commonly used SIRT algorithm (Figure 6C). This is not surprising since it is suitable for Poisson distributed data typical in STEM imaging, whereas SIRT is based on normally distributed data [64]. EM was even more outperformed by TVM with a remarkably low resulting shape error of 0.58% for a tilt range of -77° to +72°. This is expected, since TVM incorporates additional prior knowledge about the smoothness of the reconstructed object, which allows for compensating the noise and limited angular sampling range artefacts. The same conclusion was drawn for more complex reconstructed object shapes, such as a triangle and an octopod (Figure 6D). The triangle served as an example of a shape that is more susceptible to the limited angular sampling range artefacts, which stems from the alignment of the particle with respect to the tilt axis. Whereas for elongated shapes like nanorods, the missing information can be reduced by positioning the nanorod perpendicular to the tilt axis as done here, a triangle cannot be rotated in a similar optimal manner. An octopod, on the other hand, is an example of a shape with smaller and sharper geometrical features. For both of these more challenging shapes, utilizing TVM led to the reconstructions with the smallest shape error similar to the nanorod case.

Figure 6D also demonstrates that the choice of segmentation method becomes crucial when the limited angular sample range produces larger artefacts as is the case for the triangle. In the case of the nanorod or the octopod, using the Minimum or Otsu segmentation method resulted in similar shape errors, although the Minimum method slightly outperformed the Otsu one for both the TVM and EM reconstructions. However, for the nanotriangle the segmentation method had a significant influence. Using the Minimum threshold almost tripled the shape error compared to the Otsu method for the TVM and EM reconstructions. The reason behind this is detailed in section S2.2.3 in the SI. In short, the Minimum method used here calculates the minimum in the smoothed intensity histogram, which is much more sensitive to noise in the reconstruction and therefore produces less predictable segmentation results. The Otsu method, on the other hand, minimizes the inter-class variance, which is significantly more robust in the case of noisier and lower quality data. Hence, the Minimum method can be assumed to work less well for noisier data, which includes shapes that suffer from a larger influence due to a limited angular range, and should be applied to high signal-to-noise data only. It is advisable to look at the actual histograms to help with the judgement (see Figure S4).

The 3D visualizations of the final segmented TVM reconstructions of the three simulated particle shapes are displayed in Figure 7A-C. The high quality of the reconstruction and segmentation as evidenced by the low shape errors is clearly visually reproduced. Figure 7D-I demonstrates why TVM led to a smaller shape error, in particular for the triangle and the octopod. Representative slices of reconstructions using either EM (Figure 7D-F) or TVM (Figure 7G-I) for the three different shapes are shown. Strikingly, the TVM reconstructions had a significantly higher subjective quality than the EM reconstructions, as they were less noisy and displayed a less significant effect of the limited angular sampling range. This is also reflected in the voxel intensity histograms, which displayed clearer separation of the foreground and background compared to EM (Figure S4). Consequently, segmentation (segmented boundaries are displayed in light blue) was easier and more robust on the TVM data. As a result, the quantitative shape errors obtained for the TVM reconstructions were surprisingly low even for the more challenging shapes.

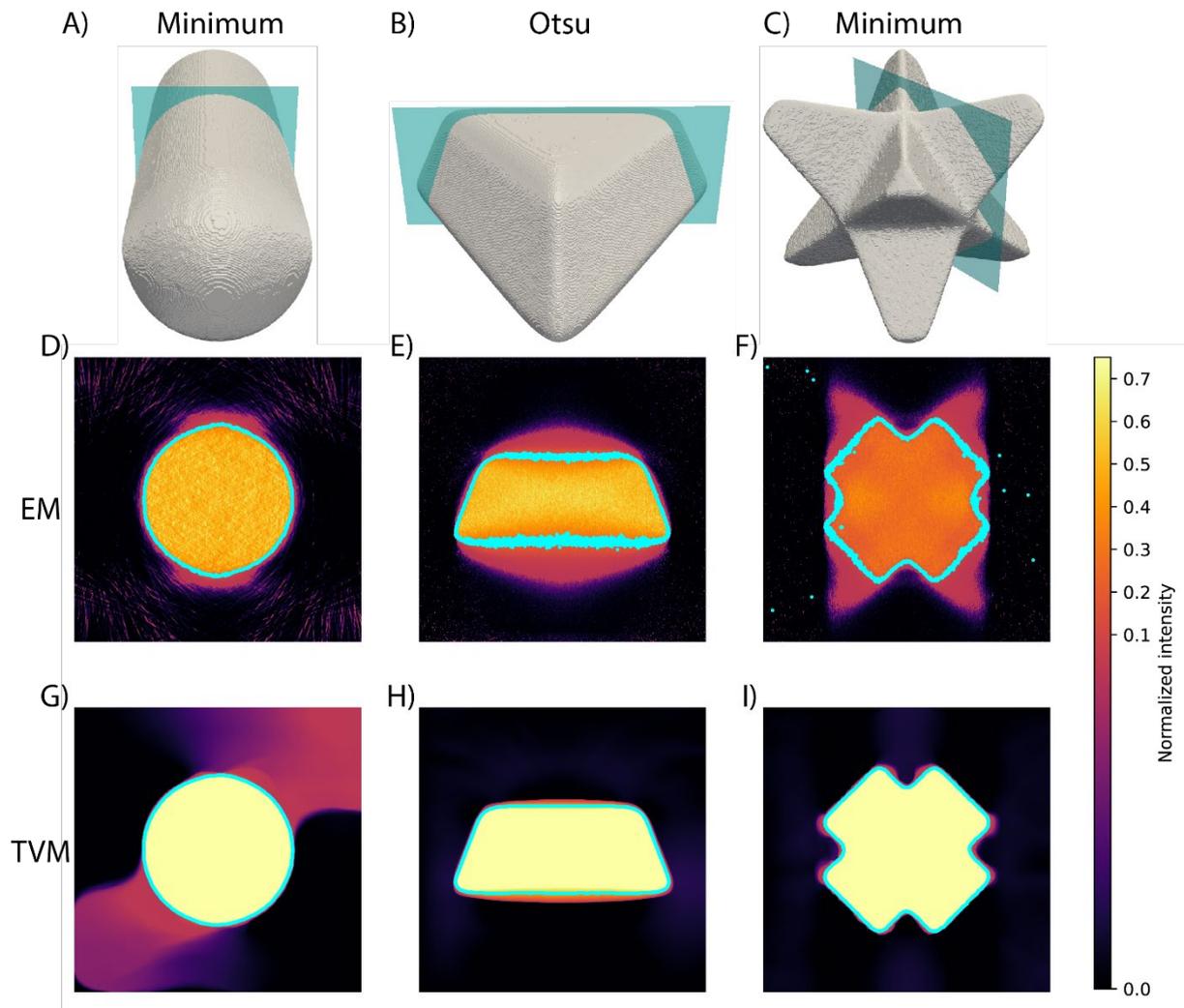

**Figure 7.** 3D visualizations of the (A) nanorod reconstructed with TVM and segmented with the Minimum threshold, (B) triangle reconstructed with TVM and segmented with Otsu threshold and (C) the octopod reconstructed with TVM and segmented with the Minimum threshold. Slices through the (D-F) EM and (G-I) TVM reconstructions before segmentation highlight the differences in the reconstruction methods. The slices were taken along the blue planes in (A-C). The light blue outlines in (D-I) correspond to the segmentations mentioned above and the double-sloped colormap is used to aid in visualizing both background and foreground noise in these segmentations.

To proceed with electromagnetic simulations based on the BEM method, the obtained reconstructions need to be converted to surface meshes. One possible approach is to fit a 3D model of a particle to the reconstruction data as we did for the experimental data to obtain the length and diameter for our ground truth simulations (Figure S6). However, this introduces an additional shape error because it is just an approximation of the shape. In fact, for the experimentally measured nanorod in Figure 4, the discrepancy between shape fitting and directly meshing of the particle resulted in a shape error of almost 5%. It is therefore beneficial to create the surface mesh from voxel data directly. The most popular algorithm for achieving this task is the marching cubes method [92], [93]. In this algorithm, segmented 3D data on a voxel grid are converted into a mesh by placing triangles at the boundary of the object with their orientations determined from the local arrangement of voxels in the segmentation. Surface meshing of the reconstructions did not result in significantly larger shape errors compared to Figure 6, see section S2.2.7 and Figure S8 for details. We observed small but noticeable differences in obtained shape error for different implementations of the marching cubes algorithm (see section S2.2.7 for a full discussion). It should be noted that we had to slightly smooth the reconstructions with Gaussian of pixel size 1 to be able to create surface meshes for all reconstructions presented in Figure 6. Without smoothing, some of the created meshes contained otherwise holes, which could not always be fixed. Whereas we did not see an effect of smoothing of the reconstructed 3D data set for the less noisy reconstructions (Figure S5), smoothing led to a significantly decreased shape error for the reconstructions performed by SIRT and 100 iterations of EM, which were noisier compared to the rest. In that case of a more limited angular tilt range of ±50° the actual missing information could non-surprisingly not be retrieved through smoothing (Figure S9).

The marching cubes algorithm usually produces a mesh with the same resolution as the input voxel data, which leads to a number of triangles on the order of $10^6$ in our case. In fact, the 3D visualizations in Figure 7A-C are these surface meshes. Such a large mesh size makes it computationally intractable to perform electromagnetic simulations [40]. For this reason, we used a mesh simplification algorithm that reduced the number of triangles to a user specified value [94]. After comparing several algorithms in terms of the shape error introduced by mesh simplification (see Figure S8 in the SI), we chose to use the so-called fast simplification algorithm, a quadric error metric-based algorithm, which iteratively removes mesh edges that contribute the least to the final simplification error. For this fast simplification algorithm, an aggression parameter needs to be chosen, which determines how aggressively faces are removed from the mesh. We found that 7 was a suitable aggression parameter (Figure S7).

The final test is to identify how the different processing steps influence the simulated scattering cross sections, which is displayed in Figure 8. Figure 8A displays the scattering cross sections and Figure 8B plots the corresponding spectrum errors (Equation 2) as a function of the shape errors with respect to the voxelized ground truth for the nanorod (same as reported in Figure 6). The ground truth for the spectrum error was based on a spherocylinder mesh with the dimensions of 30 nm x 96.5 nm and optimal discretization (see Figure 2), corresponding to 4960 triangles. For a direct comparison, all other meshes were simplified to the same number of triangles. Note that because of mesh simplification, even the ground truth model for spectrum error has a shape error of about 1%.

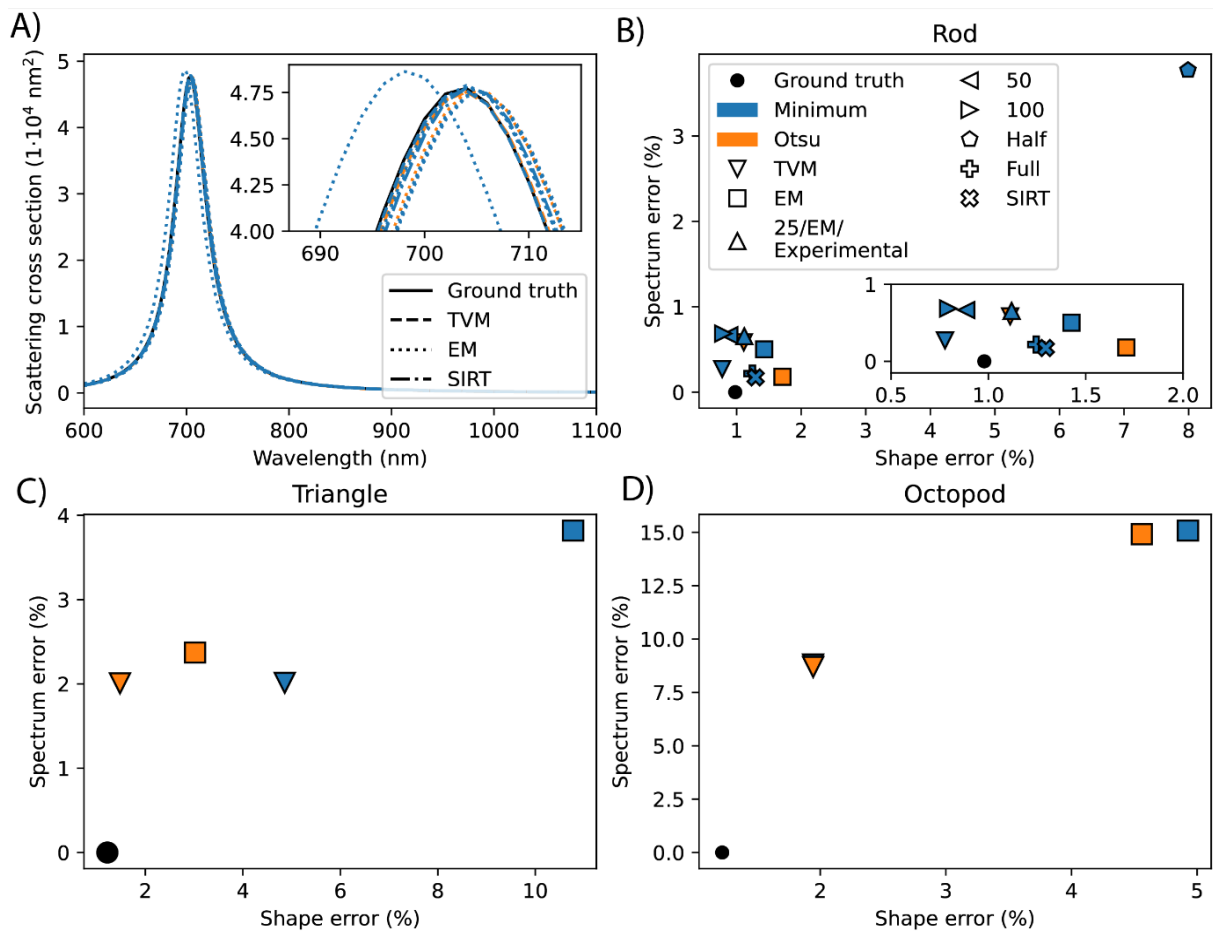

**Figure 8.** The effect of the shape error on the spectrum error. (A) Simulated BEM scattering spectra of meshes created by reconstructing, segmenting, smoothing, meshing, and simplifying simulated HAADF-STEM data of the rod. (B) Their corresponding shape and spectrum errors. (C) The shape and spectrum errors for the triangle meshes and (D) for the octopod meshes. It should be noted that all results from Figure 6 are included, which means that there are two different thresholded results for TVM (dashed line and triangle markers) and EM (dotted line and square markers), i.e. using the Minimum (blue) and Otsu (orange) methods. The numbered labels correspond to results that were reconstructed using a different number of EM iterations. 'Half' and 'Full' correspond to the used angular sampling range for the reconstruction. The labels '50' and '100' also correspond to the number of iterations for the EM reconstruction using the experimental angular range, which was the same as for 25 iterations, labelled here '25/EM/Experimental'.

We first compared the simplified surface meshes based on the segmented reconstructions obtained with different tilt ranges, reconstruction algorithms and segmentation methods from Figure 6 for the nanorod shape. Both spectrum and shape errors for the majority of cases were very low, below 1% and 2%, respectively. This is because a nanorod is a simple, symmetric shape and different investigated data processing steps, such as reconstruction smoothing and mesh simplification, are effective in removing artifacts originating from noise and suboptimal reconstruction parameters. In turn, the remaining small shape deviations do not significantly influence the spectral response, and there is no clear correlation between the shape and spectral error in this regime. In contrast to the data processing parameters, limited input data, as in the case of strongly restricted angular range reconstruction (pentagon symbol in Figure 8B), led to significant shape and spectrum errors.

The fact that the spectrum errors for the different reconstruction and data processing parameters were mainly below 1% with our workflow demonstrates that our meshing pipeline is rather robust and can create low spectrum errors even in the case of sub-optimal reconstruction choices. However, from Figure 6D we know that the nanorod is actually the most forgiving shape in terms of reconstruction and segmentation workflow. The situation is indeed different for the more challenging shapes of the triangle and octopod shown in Figure 8C and D (with their corresponding scattering spectra in Figure S10A,B). In both cases we meshed the voxelized ground truth from Figure 7B,C to use as the reference for the spectrum error. Same as for the nanorod shape, mesh simplification led to a small shape error of below 1%. For the ground truth mesh two general observations can be made for these more complex shapes. First, the higher shape errors compared to the nanorod shape led to significantly higher spectrum errors. Second, even with comparable shape errors, the spectrum errors were significantly higher for shapes with higher complexity. Whereas a 5% shape error for the nanotriangle still resulted in a spectrum error around 2%, for the octopod the spectrum error increased to 15% for a similar shape error. The reason is that more important morphological features are affected by the missing shape information. For triangles, the reconstructed shape inaccuracy mainly resulted in thickness variations (Figure 7E). For octopods, a higher shape error was connected to blunting of the tips, which blue-shifted and decreased the scattering cross section (Figure S10B). Thus, the more complex the shape, the better the reconstruction needs to be for a successful electromagnetic simulation. In our comparison, TVM performed significantly better than other algorithms because of incorporating additional prior knowledge about the reconstructed object. A promising future direction could be employing reconstruction methods based on mesh representation [95], which would allow for minimizing shape errors stemming from mesh simplification.

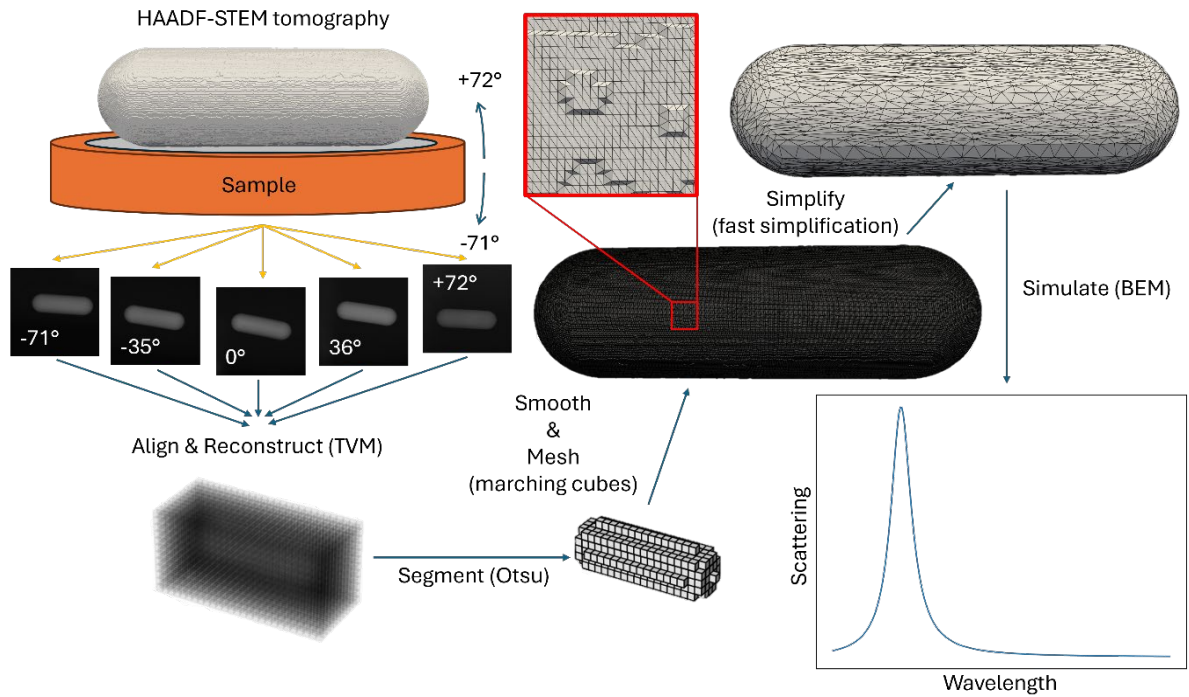

**Figure 9.** The overall proposed workflow in which HAADF-STEM tomography is performed on a nanorod. The resulting projections are aligned and reconstructed with TVM. These voxelized data are segmented with the Otsu method and smoothed before meshing with marching cubes. The resulting mesh is simplified using the fast simplification algorithm and the result is used as input for a simulation with BEM.

The overall proposed workflow and main findings are summarized in Figure 9. For optimal results, and in particular for complex shapes, we recommend to use TVM as a reconstruction algorithm together with Otsu segmentation, which proved to be more generally robust compared to other segmentation methods. To transform the segmented reconstruction into a surface mesh, the marching cubes algorithm worked well for all shapes analyzed here. We recommend to smooth the segmented reconstruction with 1 px before meshing to create a watertight mesh. We further recommend to use the fast simplification algorithm with an aggression parameter of 7 to reduce the number of surface elements. This is needed to ensure that the electromagnetic simulations can be performed in a feasible time. For the simulations itself, we found that BEM performed the best for our purpose in terms of accuracy and speed. Finally, special attention should be given to the dielectric function and accurate description of the local dielectric surrounding.

## Conclusion

In conclusion, performing electromagnetic simulations of plasmonic nanoparticles is an intricate interplay between different factors that play a role. In this work, we quantified possible error sources for a simulation workflow taking gold nanoparticles as an example system. First, we identified that BEM was a reliable simulation method with a clear convergence behavior and orders of magnitude faster simulation times compared to other conventional methods. Second, we demonstrated that even supposedly less important meshing parameters can be critical in the accuracy of the simulations and that the meshing accuracy needs to be tuned more thoroughly as is normally done. In addition, the choice of metal dielectric function ideally reflects the experimental system as it has a significant influence on the simulated optical cross sections and for accurate results, the substrate needs to be included as well. Third, BEM is known for its rather complex parametrization as it requires a triangular surface mesh as input, which is often seen as a hindrance for using it for complex morphologies. We demonstrated that using morphologies obtained from electron tomography can circumvent that problem and we developed an optimal workflow to turn a voxel-based reconstruction into a surface mesh by quantifying the introduced shape errors for different steps. Although for volume-based simulation methods the voxelized tomography output can be directly used as an input for the simulations, the less predictable convergence behaviour might not be favourable. In the end, turning the reconstruction into a surface mesh to be able to use BEM can be completely automated when following our steps. In terms of reconstruction algorithm, for all nanoparticle shapes, TVM significantly outperformed EM and SIRT. The optimal segmentation method depended on the nanoparticle shape. In general, the Otsu method was more robust and is likely the best method for single nanoparticle shapes like the ones studied here. However, for high quality and low noise data, the Minimum method performed slightly better although it is more difficult to evaluate its performance without knowing the ground truth. We demonstrated that these different processing steps can alter the final input morphology, which can in turn result in errors when simulating the optical response. Although slight smoothing of the reconstruction and the necessary surface mesh simplification could additionally lower the shape error of the object, we observed that the best approach is to enforce object smoothness during the reconstruction process rather than before. We also observed that the same shape error did not translate into a similar spectrum error for the different nanoparticle shapes, in particular when high curvature features are affected by the shape inaccuracies. The discussed topics in our work can help to achieve more accurate simulations and therefore bridge the gap between experimental optical cross sections and simulated ones by minimizing artificial discrepancies stemming from sub-optimal morphology retrieval, and thereby possibly allowing for a more accurate retrieval of the nanoparticle morphology from optical data alone. Similar considerations are valid for correlation of electron-based spectroscopies and electron tomography data and our workflow can be applied in that case as well.

# Author statements

**Acknowledgement:** We thank Tristan van Leeuwen for discussions about this work.

**Research funding:** W.A. acknowledges financial support from the research program of AMOLF, which is partly financed by the Dutch Research Council (NWO). S.B. acknowledges financial support from the European Commission under the Horizon 2020 Programme by ERC Consolidator grant no. 815128 (REALNANO).

**Author contributions:** W. A. initiated the project. M. D. performed the research, analyzed the data and wrote the original draft of the manuscript. W. A. and A. S. edited the draft. N. C. performed the electron tomography experiments under the supervision of S. B. A. S. performed the TVM reconstructions and helped with the data interpretation. All authors have accepted responsibility for the entire content of this manuscript and approved its submission.

**Conflict of interest:** Authors state no conflict of interest.

**Data availability statement:** The datasets generated during and/or analyzed during the current study will be made available on ZENODO after the revision process.

[1]   L. Wang, M. Hasanzadeh Kafshgari, and M. Meunier, "Optical Properties and Applications of Plasmonic-Metal Nanoparticles," *Adv Funct Mater*, vol. 30, no. 51, p. 2005400, Dec. 2020, doi: 10.1002/adfm.202005400.

[2]   P. Zijlstra, J. W. M. Chon, and M. Gu, "Five-dimensional optical recording mediated by surface plasmons in gold nanorods," *Nature*, vol. 459, no. 7245, pp. 410–413, May 2009, doi: 10.1038/nature08053.

[3]   W. Li *et al.*, "Bidirectional plasmonic coloration with gold nanoparticles by wavelength-switched photoredox reaction," *Nanoscale*, vol. 10, no. 46, pp. 21910–21917, 2018, doi: 10.1039/C8NR05763J.

[4]   J. N. Anker, W. P. Hall, O. Lyandres, N. C. Shah, J. Zhao, and R. P. Van Duyne, "Biosensing with plasmonic nanosensors," *Nat Mater*, vol. 7, no. 6, pp. 442–453, Jun. 2008, doi: 10.1038/nmat2162.

[5]   B. S. Hoener *et al.*, "Plasmonic Sensing and Control of Single-Nanoparticle Electrochemistry," *Chem*, vol. 4, no. 7, pp. 1560–1585, Jul. 2018, doi: 10.1016/j.chempr.2018.04.009.

[6]   J. B. Sambur and P. Chen, "Approaches to Single-Nanoparticle Catalysis," *Annu Rev Phys Chem*, vol. 65, no. 1, pp. 395–422, Apr. 2014, doi: 10.1146/annurev-physchem-040513-103729.

[7] S. Yu, A. J. Wilson, G. Kumari, X. Zhang, and P. K. Jain, "Opportunities and Challenges of Solar-Energy-Driven Carbon Dioxide to Fuel Conversion with Plasmonic Catalysts," *ACS Energy Lett*, vol. 2, no. 9, pp. 2058–2070, Sep. 2017, doi: 10.1021/acsenergylett.7b00640.

[8] K. Sytwu, M. Vadai, and J. A. Dionne, "Bimetallic nanostructures: combining plasmonic and catalytic metals for photocatalysis," *Adv Phys X*, vol. 4, no. 1, p. 1619480, Jan. 2019, doi: 10.1080/23746149.2019.1619480.

[9] S. Alekseeva, I. I. Nedrygailov, and C. Langhammer, "Single Particle Plasmonics for Materials Science and Single Particle Catalysis," *ACS Photonics*, vol. 6, no. 6, pp. 1319–1330, Jun. 2019, doi: 10.1021/acsphotonics.9b00339.

[10] C. C. Carlin *et al.*, "Nanoscale and ultrafast *in situ* techniques to probe plasmon photocatalysis," *Chemical Physics Reviews*, vol. 4, no. 4, Dec. 2023, doi: 10.1063/5.0163354.

[11] A. Baldi and S. H. C. Askes, "Pulsed Photothermal Heterogeneous Catalysis," *ACS Catal*, vol. 13, no. 5, pp. 3419–3432, Mar. 2023, doi: 10.1021/acscatal.2c05435.

[12] K. L. Kelly, E. Coronado, L. L. Zhao, and G. C. Schatz, "The Optical Properties of Metal Nanoparticles:  The Influence of Size, Shape, and Dielectric Environment," *J Phys Chem B*, vol. 107, no. 3, pp. 668–677, Jan. 2003, doi: 10.1021/jp026731y.

[13] L. Scarabelli, A. Sánchez-Iglesias, J. Pérez-Juste, and L. M. Liz-Marzán, "A 'Tips and Tricks' Practical Guide to the Synthesis of Gold Nanorods," *J Phys Chem Lett*, vol. 6, no. 21, pp. 4270–4279, Nov. 2015, doi: 10.1021/acs.jpclett.5b02123.

[14] L. M. Liz-Marzán, C. R. Kagan, and J. E. Millstone, "Reproducibility in Nanocrystal Synthesis? Watch Out for Impurities!," *ACS Nano*, vol. 14, no. 6, pp. 6359–6361, Jun. 2020, doi: 10.1021/acsnano.0c04709.

[15] J. E. Ortiz-Castillo, R. C. Gallo-Villanueva, M. J. Madou, and V. H. Perez-Gonzalez, "Anisotropic gold nanoparticles: A survey of recent synthetic methodologies," *Coord Chem Rev*, vol. 425, p. 213489, Dec. 2020, doi: 10.1016/j.ccr.2020.213489.

[16] S. E. Lohse, N. D. Burrows, L. Scarabelli, L. M. Liz-Marzán, and C. J. Murphy, "Anisotropic Noble Metal Nanocrystal Growth: The Role of Halides," *Chemistry of Materials*, vol. 26, no. 1, pp. 34–43, Jan. 2014, doi: 10.1021/cm402384j.

[17] G. Lin, W. Lu, W. Cui, and L. Jiang, "A Simple Synthesis Method for Gold Nano- and Microplate Fabrication Using a Tree-Type Multiple-Amine Head Surfactant," *Cryst Growth Des*, vol. 10, no. 3, pp. 1118–1123, Mar. 2010, doi: 10.1021/cg9008976.

[18] F. Kim, S. Connor, H. Song, T. Kuykendall, and P. Yang, "Platonic Gold Nanocrystals," *Angewandte Chemie International Edition*, vol. 43, no. 28, pp. 3673–3677, Jul. 2004, doi: 10.1002/anie.200454216.

[19] S. Barbosa *et al.*, "Tuning Size and Sensing Properties in Colloidal Gold Nanostars," *Langmuir*, vol. 26, no. 18, pp. 14943–14950, Sep. 2010, doi: 10.1021/la102559e.

[20] A. Guerrero-Martínez, S. Barbosa, I. Pastoriza-Santos, and L. M. Liz-Marzán, "Nanostars shine bright for you," *Curr Opin Colloid Interface Sci*, vol. 16, no. 2, pp. 118–127, Apr. 2011, doi: 10.1016/j.cocis.2010.12.007.

[21]  S. Adhikari, P. Spaeth, A. Kar, M. D. Baaske, S. Khatua, and M. Orrit, "Photothermal Microscopy: Imaging the Optical Absorption of Single Nanoparticles and Single Molecules," *ACS Nano*, vol. 14, no. 12, pp. 16414–16445, Dec. 2020, doi: 10.1021/acsnano.0c07638.

[22]  B. Ni *et al.*, "Chiral Seeded Growth of Gold Nanorods Into Fourfold Twisted Nanoparticles with Plasmonic Optical Activity," *Advanced Materials*, vol. 35, no. 1, Jan. 2023, doi: 10.1002/adma.202208299.

[23]  K. Van Gordon *et al.*, "Tuning the Growth of Chiral Gold Nanoparticles Through Rational Design of a Chiral Molecular Inducer," *Nano Lett*, vol. 23, no. 21, pp. 9880–9886, Nov. 2023, doi: 10.1021/acs.nanolett.3c02800.

[24]  G. González-Rubio *et al.*, "Micelle-directed chiral seeded growth on anisotropic gold nanocrystals," *Science (1979)*, vol. 368, no. 6498, pp. 1472–1477, Jun. 2020, doi: 10.1126/science.aba0980.

[25]  J. Rodríguez-Fernández *et al.*, "Spectroscopy, Imaging, and Modeling of Individual Gold Decahedra," *The Journal of Physical Chemistry C*, vol. 113, no. 43, pp. 18623–18631, Oct. 2009, doi: 10.1021/jp907646d.

[26]  E. Ringe, B. Sharma, A.-I. Henry, L. D. Marks, and R. P. Van Duyne, "Single nanoparticle plasmonics," *Physical Chemistry Chemical Physics*, vol. 15, no. 12, p. 4110, 2013, doi: 10.1039/c3cp44574g.

[27]  R. F. Hamans, R. Kamarudheen, and A. Baldi, "Single Particle Approaches to Plasmon-Driven Catalysis," *Nanomaterials*, vol. 10, no. 12, p. 2377, Nov. 2020, doi: 10.3390/nano10122377.

[28]  P. Zijlstra and M. Orrit, "Single metal nanoparticles: optical detection, spectroscopy and applications," *Reports on Progress in Physics*, vol. 74, no. 10, p. 106401, Oct. 2011, doi: 10.1088/0034-4885/74/10/106401.

[29]  J. Olson, S. Dominguez-Medina, A. Hoggard, L.-Y. Wang, W.-S. Chang, and S. Link, "Optical characterization of single plasmonic nanoparticles," *Chem Soc Rev*, vol. 44, no. 1, pp. 40–57, 2015, doi: 10.1039/C4CS00131A.

[30]  A.-I. Henry, J. M. Bingham, E. Ringe, L. D. Marks, G. C. Schatz, and R. P. Van Duyne, "Correlated Structure and Optical Property Studies of Plasmonic Nanoparticles," *The Journal of Physical Chemistry C*, vol. 115, no. 19, pp. 9291–9305, May 2011, doi: 10.1021/jp2010309.

[31]  T. Jollans, M. D. Baaske, and M. Orrit, "Nonfluorescent Optical Probing of Single Molecules and Nanoparticles," *The Journal of Physical Chemistry C*, vol. 123, no. 23, pp. 14107–14117, Jun. 2019, doi: 10.1021/acs.jpcc.9b00843.

[32]  A. Al-Zubeidi, L. A. McCarthy, A. Rafiei-Miandashti, T. S. Heiderscheit, and S. Link, "Single-particle scattering spectroscopy: fundamentals and applications," *Nanophotonics*, vol. 10, no. 6, pp. 1621–1655, Apr. 2021, doi: 10.1515/nanoph-2020-0639.

[33]  M. Dieperink, F. Scalerandi, and W. Albrecht, "Correlating structure, morphology and properties of metal nanostructures by combining single-particle optical spectroscopy

and electron microscopy," *Nanoscale*, vol. 14, no. 20, pp. 7460–7472, 2022, doi: 10.1039/D1NR08130F.

[34] E. M. Perassi *et al.*, "Quantitative Understanding of the Optical Properties of a Single, Complex-Shaped Gold Nanoparticle from Experiment and Theory," *ACS Nano*, vol. 8, no. 5, pp. 4395–4402, May 2014, doi: 10.1021/nn406270z.

[35] V. Myroshnychenko, E. Carbó-Argibay, I. Pastoriza-Santos, J. Pérez-Juste, L. M. Liz-Marzán, and F. J. García de Abajo, "Modeling the Optical Response of Highly Faceted Metal Nanoparticles with a Fully 3D Boundary Element Method," *Advanced Materials*, vol. 20, no. 22, pp. 4288–4293, Nov. 2008, doi: 10.1002/adma.200703214.

[36] A. Zilli, W. Langbein, and P. Borri, "Quantitative Measurement of the Optical Cross Sections of Single Nano-objects by Correlative Transmission and Scattering Microspectroscopy," *ACS Photonics*, vol. 6, no. 8, pp. 2149–2160, Aug. 2019, doi: 10.1021/acsphotonics.9b00727.

[37] Y. Wang, A. Zilli, Z. Sztranyovszky, W. Langbein, and P. Borri, "Quantitative optical microspectroscopy, electron microscopy, and modelling of individual silver nanocubes reveal surface compositional changes at the nanoscale," *Nanoscale Adv*, vol. 2, no. 6, pp. 2485–2496, 2020, doi: 10.1039/D0NA00059K.

[38] L. M. Payne, W. Albrecht, W. Langbein, and P. Borri, "The optical nanosizer – quantitative size and shape analysis of individual nanoparticles by high-throughput widefield extinction microscopy," *Nanoscale*, vol. 12, no. 30, pp. 16215–16228, 2020, doi: 10.1039/D0NR03504A.

[39] L. M. Payne, F. Masia, A. Zilli, W. Albrecht, P. Borri, and W. Langbein, "Quantitative morphometric analysis of single gold nanoparticles by optical extinction microscopy: Material permittivity and surface damping effects," *J Chem Phys*, vol. 154, no. 4, Jan. 2021, doi: 10.1063/5.0031012.

[40] Y. Wang *et al.*, "Quantitatively linking morphology and optical response of individual silver nanohedra," *Nanoscale*, vol. 14, no. 30, pp. 11028–11037, 2022, doi: 10.1039/D2NR02131E.

[41] K. Jenkinson, L. M. Liz-Marzán, and S. Bals, "Multimode Electron Tomography Sheds Light on Synthesis, Structure, and Properties of Complex Metal-Based Nanoparticles," *Advanced Materials*, vol. 34, no. 36, Sep. 2022, doi: 10.1002/adma.202110394.

[42] W. Albrecht, S. Van Aert, and S. Bals, "Three-Dimensional Nanoparticle Transformations Captured by an Electron Microscope," *Acc Chem Res*, vol. 54, no. 5, pp. 1189–1199, Mar. 2021, doi: 10.1021/acs.accounts.0c00711.

[43] A. Skorikov *et al.*, "Quantitative 3D Characterization of Elemental Diffusion Dynamics in Individual Ag@Au Nanoparticles with Different Shapes," *ACS Nano*, vol. 13, no. 11, pp. 13421–13429, Nov. 2019, doi: 10.1021/acsnano.9b06848.

[44] T. Milagres de Oliveira *et al.*, "3D Characterization and Plasmon Mapping of Gold Nanorods Welded by Femtosecond Laser Irradiation," *ACS Nano*, vol. 14, no. 10, pp. 12558–12570, Oct. 2020, doi: 10.1021/acsnano.0c02610.

[45]	W. Albrecht, E. Bladt, H. Vanrompay, J. D. Smith, S. E. Skrabalak, and S. Bals, "Thermal Stability of Gold/Palladium Octopods Studied *in Situ* in 3D: Understanding Design Rules for Thermally Stable Metal Nanoparticles," *ACS Nano*, vol. 13, no. 6, pp. 6522–6530, Jun. 2019, doi: 10.1021/acsnano.9b00108.

[46]	H. Vanrompay *et al.*, "3D characterization of heat-induced morphological changes of Au nanostars by fast *in situ* electron tomography," *Nanoscale*, vol. 10, no. 48, pp. 22792–22801, 2018, doi: 10.1039/C8NR08376B.

[47]	Y. Wang, A. B. Serrano, K. Sentosun, S. Bals, and L. M. Liz-Marzán, "Stabilization and Encapsulation of Gold Nanostars Mediated by Dithiols," *Small*, vol. 11, no. 34, pp. 4314–4320, Sep. 2015, doi: 10.1002/smll.201500703.

[48]	P. C. Angelomé *et al.*, "Seedless Synthesis of Single Crystalline Au Nanoparticles with Unusual Shapes and Tunable LSPR in the near-IR," *Chemistry of Materials*, vol. 24, no. 7, pp. 1393–1399, Apr. 2012, doi: 10.1021/cm3004479.

[49]	P. A. Midgley and M. Weyland, "3D electron microscopy in the physical sciences: the development of Z-contrast and EFTEM tomography," *Ultramicroscopy*, vol. 96, no. 3–4, pp. 413–431, Sep. 2003, doi: 10.1016/S0304-3991(03)00105-0.

[50]	W. Albrecht and S. Bals, "Fast Electron Tomography for Nanomaterials," *The Journal of Physical Chemistry C*, vol. 124, no. 50, pp. 27276–27286, Dec. 2020, doi: 10.1021/acs.jpcc.0c08939.

[51]	P. Spaeth *et al.*, "Photothermal Circular Dichroism Measurements of Single Chiral Gold Nanoparticles Correlated with Electron Tomography," *ACS Photonics*, vol. 9, no. 12, pp. 3995–4004, Dec. 2022, doi: 10.1021/acsphotonics.2c01457.

[52]	M. Li, S. K. Cushing, and N. Wu, "Plasmon-enhanced optical sensors: a review," *Analyst*, vol. 140, no. 2, pp. 386–406, 2015, doi: 10.1039/C4AN01079E.

[53]	E. Cortés *et al.*, "Challenges in Plasmonic Catalysis," *ACS Nano*, vol. 14, no. 12, pp. 16202–16219, Dec. 2020, doi: 10.1021/acsnano.0c08773.

[54]	K. Shiratori *et al.*, "Machine-Learned Decision Trees for Predicting Gold Nanorod Sizes from Spectra," *The Journal of Physical Chemistry C*, vol. 125, no. 35, pp. 19353–19361, Sep. 2021, doi: 10.1021/acs.jpcc.1c03937.

[55]	J. S. Googasian and S. E. Skrabalak, "Practical Considerations for Simulating the Plasmonic Properties of Metal Nanoparticles," *ACS Physical Chemistry Au*, vol. 3, no. 3, pp. 252–262, May 2023, doi: 10.1021/acsphyschemau.2c00064.

[56]	A. Trügler, *Optical Properties of Metallic Nanoparticles*, vol. 232. Cham: Springer International Publishing, 2016. doi: 10.1007/978-3-319-25074-8.

[57]	A. Amirjani and S. K. Sadrnezhaad, "Computational electromagnetics in plasmonic nanostructures," Aug. 21, 2021, *Royal Society of Chemistry*. doi: 10.1039/d1tc01742j.

[58]	M. R. Gonçalves, "Plasmonic nanoparticles: fabrication, simulation and experiments," *J Phys D Appl Phys*, vol. 47, no. 21, p. 213001, May 2014, doi: 10.1088/0022-3727/47/21/213001.

[59] E. Ringe *et al.*, "Unraveling the Effects of Size, Composition, and Substrate on the Localized Surface Plasmon Resonance Frequencies of Gold and Silver Nanocubes: A Systematic Single-Particle Approach," *The Journal of Physical Chemistry C*, vol. 114, no. 29, pp. 12511–12516, Jul. 2010, doi: 10.1021/jp104366r.

[60] J. Rodríguez-Fernández, I. Pastoriza-Santos, J. Pérez-Juste, F. J. García de Abajo, and L. M. Liz-Marzán, "The Effect of Silica Coating on the Optical Response of Sub-micrometer Gold Spheres," *The Journal of Physical Chemistry C*, vol. 111, no. 36, pp. 13361–13366, Sep. 2007, doi: 10.1021/jp073853n.

[61] R. E. Armstrong, J. C. van Liempt, and P. Zijlstra, "Effect of Film Thickness on the Far- and Near-Field Optical Response of Nanoparticle-on-Film Systems," *The Journal of Physical Chemistry C*, vol. 123, no. 42, pp. 25801–25808, Oct. 2019, doi: 10.1021/acs.jpcc.9b06592.

[62] E. M. Perassi, Juan. C. Hernandez-Garrido, M. S. Moreno, E. R. Encina, E. A. Coronado, and P. A. Midgley, "Using Highly Accurate 3D Nanometrology to Model the Optical Properties of Highly Irregular Nanoparticles: A Powerful Tool for Rational Design of Plasmonic Devices," *Nano Lett*, vol. 10, no. 6, pp. 2097–2104, Jun. 2010, doi: 10.1021/nl1005492.

[63] J. Marcheselli *et al.*, "Simulating Plasmon Resonances of Gold Nanoparticles with Bipyramidal Shapes by Boundary Element Methods," *J Chem Theory Comput*, vol. 16, no. 6, pp. 3807–3815, Jun. 2020, doi: 10.1021/acs.jctc.0c00269.

[64] B. Goris, T. Roelandts, K. J. Batenburg, H. Heidari Mezerji, and S. Bals, "Advanced reconstruction algorithms for electron tomography: From comparison to combination," *Ultramicroscopy*, vol. 127, pp. 40–47, Apr. 2013, doi: 10.1016/j.ultramic.2012.07.003.

[65] C. A. Glasbey, "An Analysis of Histogram-Based Thresholding Algorithms," *CVGIP: Graphical Models and Image Processing*, vol. 55, no. 6, pp. 532–537, Nov. 1993, doi: 10.1006/cgip.1993.1040.

[66] T. MacDonald, "Isosurface Extraction," swiftcoding. Accessed: Apr. 28, 2024. [Online]. Available: https://swiftcoder.wordpress.com/planets/isosurface-extraction/

[67] G. Mie, "Beiträge zur Optik trüber Medien, speziell kolloidaler Metallösungen," *Ann Phys*, vol. 330, no. 3, pp. 377–445, Jan. 1908, doi: 10.1002/andp.19083300302.

[68] Ansys, "Convergence testing process for FDTD simulations." Accessed: Apr. 29, 2024. [Online]. Available: https://optics.ansys.com/hc/en-us/articles/360034915833-Convergence-testing-process-for-FDTD-simulations

[69] J. Zheng *et al.*, "Gold Nanorods: The Most Versatile Plasmonic Nanoparticles," *Chem Rev*, vol. 121, no. 21, pp. 13342–13453, Nov. 2021, doi: 10.1021/acs.chemrev.1c00422.

[70] U. Hohenester and A. Trügler, "MNPBEM - A Matlab toolbox for the simulation of plasmonic nanoparticles," Sep. 2011, doi: 10.1016/j.cpc.2011.09.009.

[71] H. Sun, D. L. Darmofal, and R. Haimes, "On the impact of triangle shapes for boundary layer problems using high-order finite element discretization," *J Comput Phys*, vol. 231, no. 2, pp. 541–557, Jan. 2012, doi: 10.1016/j.jcp.2011.09.018.

[72] P. M. Knupp, "Achieving finite element mesh quality via optimization of the Jacobian matrix norm and associated quantities. Part I?a framework for surface mesh optimization," *Int J Numer Methods Eng*, vol. 48, no. 3, pp. 401–420, May 2000, doi: 10.1002/(SICI)1097-0207(20000530)48:3<401::AID-NME880>3.0.CO;2-D.

[73] R. L. Olmon *et al.*, "Optical dielectric function of gold," *Phys Rev B*, vol. 86, no. 23, p. 235147, Dec. 2012, doi: 10.1103/PhysRevB.86.235147.

[74] P. B. Johnson and R. W. Christy, "Optical Constants of the Noble Metals," *Phys Rev B*, vol. 6, no. 12, pp. 4370–4379, Dec. 1972, doi: 10.1103/PhysRevB.6.4370.

[75] E. D. Palik, *Handbook of Optical Constants of Solids*, vol. 3. Academic Press, 1998.

[76] K. M. McPeak *et al.*, "Plasmonic Films Can Easily Be Better: Rules and Recipes," *ACS Photonics*, vol. 2, no. 3, pp. 326–333, Mar. 2015, doi: 10.1021/ph5004237.

[77] Y. Lebsir, S. Boroviks, M. Thomaschewski, S. I. Bozhevolnyi, and V. A. Zenin, "Ultimate Limit for Optical Losses in Gold, Revealed by Quantitative Near-Field Microscopy," *Nano Lett*, vol. 22, no. 14, pp. 5759–5764, Jul. 2022, doi: 10.1021/acs.nanolett.2c01059.

[78] P. Stoller, V. Jacobsen, and V. Sandoghdar, "Measurement of the complex dielectric constant of a single gold nanoparticle," *Opt Lett*, vol. 31, no. 16, p. 2474, Aug. 2006, doi: 10.1364/OL.31.002474.

[79] C. Novo *et al.*, "Contributions from radiation damping and surface scattering to the linewidth of the longitudinal plasmon band of gold nanorods: a single particle study," *Physical Chemistry Chemical Physics*, vol. 8, no. 30, p. 3540, 2006, doi: 10.1039/b604856k.

[80] B. Foerster, A. Joplin, K. Kaefer, S. Celiksoy, S. Link, and C. Sönnichsen, "Chemical Interface Damping Depends on Electrons Reaching the Surface," *ACS Nano*, vol. 11, no. 3, pp. 2886–2893, Mar. 2017, doi: 10.1021/acsnano.6b08010.

[81] J. A. Scholl, A. L. Koh, and J. A. Dionne, "Quantum plasmon resonances of individual metallic nanoparticles," *Nature*, vol. 483, no. 7390, pp. 421–427, Mar. 2012, doi: 10.1038/nature10904.

[82] L. V. Rodríguez-de Marcos, J. I. Larruquert, J. A. Méndez, and J. A. Aznárez, "Self-consistent optical constants of $SiO_2$ and $Ta_2O_5$ films," *Opt Mater Express*, vol. 6, no. 11, p. 3622, Nov. 2016, doi: 10.1364/OME.6.003622.

[83] K. Luke, Y. Okawachi, M. R. E. Lamont, A. L. Gaeta, and M. Lipson, "Broadband mid-infrared frequency comb generation in a $Si_3N_4$ microresonator," *Opt Lett*, vol. 40, no. 21, p. 4823, Nov. 2015, doi: 10.1364/OL.40.004823.

[84] J. I. Larruquert, L. V. Rodríguez-de Marcos, J. A. Méndez, P. J. Martin, and A. Bendavid, "High reflectance ta-C coatings in the extreme ultraviolet," *Opt Express*, vol. 21, no. 23, p. 27537, Nov. 2013, doi: 10.1364/OE.21.027537.

[85] M. W. Knight, Y. Wu, J. B. Lassiter, P. Nordlander, and N. J. Halas, "Substrates Matter: Influence of an Adjacent Dielectric on an Individual Plasmonic Nanoparticle," *Nano Lett*, vol. 9, no. 5, pp. 2188–2192, May 2009, doi: 10.1021/nl900945q.

[86] S. Zhang, K. Bao, N. J. Halas, H. Xu, and P. Nordlander, "Substrate-Induced Fano Resonances of a Plasmonic Nanocube: A Route to Increased-Sensitivity Localized Surface Plasmon Resonance Sensors Revealed," *Nano Lett*, vol. 11, no. 4, pp. 1657–1663, Apr. 2011, doi: 10.1021/nl200135r.

[87] E. J. Kirkland, *Advanced Computing in Electron Microscopy*. Cham: Springer International Publishing, 2020. doi: 10.1007/978-3-030-33260-0.

[88] H. Vanrompay *et al.*, "Fast versus conventional HAADF-STEM tomography of nanoparticles: advantages and challenges," *Ultramicroscopy*, vol. 221, p. 113191, Feb. 2021, doi: 10.1016/j.ultramic.2020.113191.

[89] N. Otsu, "A Threshold Selection Method from Gray-Level Histograms," *IEEE Trans Syst Man Cybern*, vol. 9, no. 1, pp. 62–66, Jan. 1979, doi: 10.1109/TSMC.1979.4310076.

[90] W. van Aarle *et al.*, "The ASTRA Toolbox: A platform for advanced algorithm development in electron tomography," *Ultramicroscopy*, vol. 157, pp. 35–47, Oct. 2015, doi: 10.1016/j.ultramic.2015.05.002.

[91] H. Wieczorek, "The image quality of FBP and MLEM reconstruction," *Phys Med Biol*, vol. 55, no. 11, pp. 3161–3176, Jun. 2010, doi: 10.1088/0031-9155/55/11/012.

[92] W. E. Lorensen and H. E. Cline, "Marching cubes," in *Seminal graphics*, New York, NY, USA: ACM, 1998, pp. 347–353. doi: 10.1145/280811.281026.

[93] T. Lewiner, H. Lopes, A. W. Vieira, and G. Tavares, "Efficient Implementation of Marching Cubes' Cases with Topological Guarantees," *Journal of Graphics Tools*, vol. 8, no. 2, pp. 1–15, Jan. 2003, doi: 10.1080/10867651.2003.10487582.

[94] M. Garland and P. S. Heckbert, "Surface simplification using quadric error metrics," in *Proceedings of the 24th annual conference on Computer graphics and interactive techniques - SIGGRAPH '97*, New York, New York, USA: ACM Press, 1997, pp. 209–216. doi: 10.1145/258734.258849.

[95] J. Koo, A. B. Dahl, J. A. Bærentzen, Q. Chen, S. Bals, and V. A. Dahl, "Shape from projections via differentiable forward projector for computed tomography," *Ultramicroscopy*, vol. 224, p. 113239, May 2021, doi: 10.1016/j.ultramic.2021.113239.